# A Mini Review on The Applications of Nanomaterials in Forensic Science

Aaromal Venugopal[#], Vanshika Seth[#], Shreya Subhash Naik, Sreya Valappil, Aman Verma, Shalini Rajan, Pranav Vilas Shetgaonkar, Akshita Sinha, Sandeep Munjal

*National Forensic Sciences University, Goa, 403401, India*

**Abstract:**

Herein, we report a minireview to give a brief introduction of applications of nanomaterials in the field of forensic science. The materials that have their size in nanoscale (1 - 100 nm) comes under the category of nanomaterials. Nanomaterials possess various applications in different fields like cosmetic production, medical, photoconductivity etc. because of their physio-chemical, electrical and magnetic properties. Due to the different characteristic property that nanomaterials have, they are widely employed in diverse domains. In various fields of forensic science such as fingerprints, toxicology, medicine, serology, nanomaterials are being used extensively. Large surface area to volume ratio of the materials in nano-regime makes the nanomaterials suitable for all these application with high efficiency. This review article briefs about the nanomaterials, their advantages and their novel applications in various fields, focusing especially in the field of forensic science. The basic idea of different areas of forensic science such as development of fingerprints, detection of drugs, estimating the time since death, analysis of GSR, detection of various explosives and for the extraction of DNA etc. has also been provided.

[#] *Contributed equally.*

# 1. Introduction:

Nanomaterials have been a popular and extensively studied area for more than a decade, and this has led to the emergence of a new field of science known as 'nanotechnology' [1]. The term nano comes from a Greek word that means 'dwarf'. Nanomaterials are the materials that possess unique properties due to their size, which is typically between 1-100 nanometer, or one-millionth of a millimeter. Nanomaterials can occur naturally or can be synthesized in laboratories for specialised functions, and examples include quantum wires, ultrathin films, graphene etc. Nanomaterials are classified into different categories based on their dimensionality, chemical composition, origin and formation [2]. Richard W. Siegel classified nanomaterials into different categories; zero-dimensional (clusters), one-dimensional (nanofibers), two-dimensional (films) and three-dimensional (nanoparticles). Nanomaterials have various optical, magnetic, electrical and other properties that have a significant impact on different fields [3].

## 1.1. Advantages of nanomaterials:

Nanomaterials exhibit tremendous advantages over bulk counterparts, such as increased mechanical strength, lower melting point and decrease in surface area. According to their properties and functions, nanomaterials can be synthesised by different physical, chemical and biological methods. Fine grain size and size distribution, presence of interfaces, nature and morphology and interaction between consistent domains are the microstructural features of nanomaterials [4], [5]. The nanoscale level of nanomaterials gives them a large surface to mass ratio, making them surface dependent and suitable for various other applications in environment [3]. The quantum confinement effects of nanomaterials are related to the dimensions of the nanomaterials. The quantum mechanics and the particle wave duality cause quantum effects where the size is smaller than the material, which in turn will cause the de-Brogli wavelength of electrons and phonons to propagate [1]. The mechanical properties of nanomaterials contribute to their strength and toughness and also control defects like grain boundaries [6]. The optical properties of nanomaterials such as their size, shape and surface characteristics, make them suitable for applications in solar cells, sensors, lasers and photocatalysis. [3].

1.2. Novel applications of nanomaterials:

Nanomaterials have a wide range of applications in various fields. In the cosmetic industry, nanomaterials are incorporated with other personal care products to achieve better results such as transparency of lotion and UV protection. Carbon nanotubes are specifically used to make sports equipment to provide more strength and durability. In textile-based industries, nanomaterials also provide anti-wrinkle, water repellence and anti-microbial properties for clothing. Nanomaterials can be used to make products such as anti-graffitic coatings for walls, self-cleaning windows and crack-resistant paints which are rustproof, stronger and fire-resistant [7].

In the medical field, nanomaterials are used for drug delivery, nanomachinery (cellular mechanism and early diagnosis), tissue engineering (enhance biological functions of drugs and cells) and nanoimaging. Gold nanoparticles can be used as an antibacterial agent for bacterial infections, also the nanoparticles have applications in rapid tests for pregnancy, HIV, treatment of cancer and drug delivery [8].

The photoconductivity and wide band gap features of nanomaterials can be used in UV photo detector. Huang et al. in his paper mentioned that ZnO nanowires shows high-aspect-ratio probes for atomic force microscopy (AFM) under certain conditions. Due to its ability, a single ZnO nanowire is capable of functioning as both a nano-resonator and a nanoscale cantilever. Carbon-based composites are known for its strength, researchers are taking advantage of these nanoparticles to build nanorobots that detect damage in the body [9]. In civil engineering, nanotechnology have various properties including, light weight, better design, good construction process and stronger composites. For example, $SiO_2$ nano powders can provide better results in concrete than traditional $SiO_2$ powders. The surface to volume ratio makes it differ from the bulk materials, this feature can be used in different fields including instrumentation, energy production, electronics etc. In the area of power production field, the various applications include, rechargeable batteries by nanostructured manganese dioxide, silicon and titanium dioxide nanoparticles for thin film solar cells and fuel cells made by CNT, which can generate clean water and electricity [10].

In forensic science, nanomaterials have various applications in different fields including toxicology, biology, questioned document analysis, fingerprint detection and development. In crime scene, nanotechnology helps in the detection of trace evidences, toxins and explosives. Gold nanoparticles improves the amplification and extraction of DNA by PCR method [11].

## 2. Forensic science and applications of nanomaterials:

Different scientists defined forensic science in their own point of view. In general, forensic science is a discipline of science that involves the applied knowledge of all the other basic sciences. The Oxford English Dictionary described forensic science as a 'mixed science', where various principles and methodology of other sciences are used to serve justice to the victims and to bring the truth behind the crime into reality [12]. The birth of the term 'forensic' originated from the Latin which gave its meaning as 'forum' which means a place for public discussion. From this, definition for forensic science were inferred as "the methods and techniques of science applied to matters involving the public" [13]. The application of forensic science got wider in the last 30 years, and now they are widely used in issues related to the court of law and justice [13], [14].

In this particular domain, skills and knowledge of experts from different fields are brainstormed for smooth and effective investigation by law enforcement agencies on various cases [15]. The various sub-domains that come under the umbrella of forensic science includes Forensic Toxicology, Forensic Anthropology, Forensic Odontology, Forensic Biology, Forensic Psychiatry, Questioned Documents, Criminalistics, Jurisprudence, Fingerprints etc. The pioneers such as Mathieu Orfila, Alphonse Bertillon, Francis Galton, Albert S. Osborn, Calvin Goddard and Edmond Locard gave immense and commendable contributions for the growth of Forensic Science [15], [16]. Fig.1. gives an idea about the various novel applications of nanotechnology in forensic science.

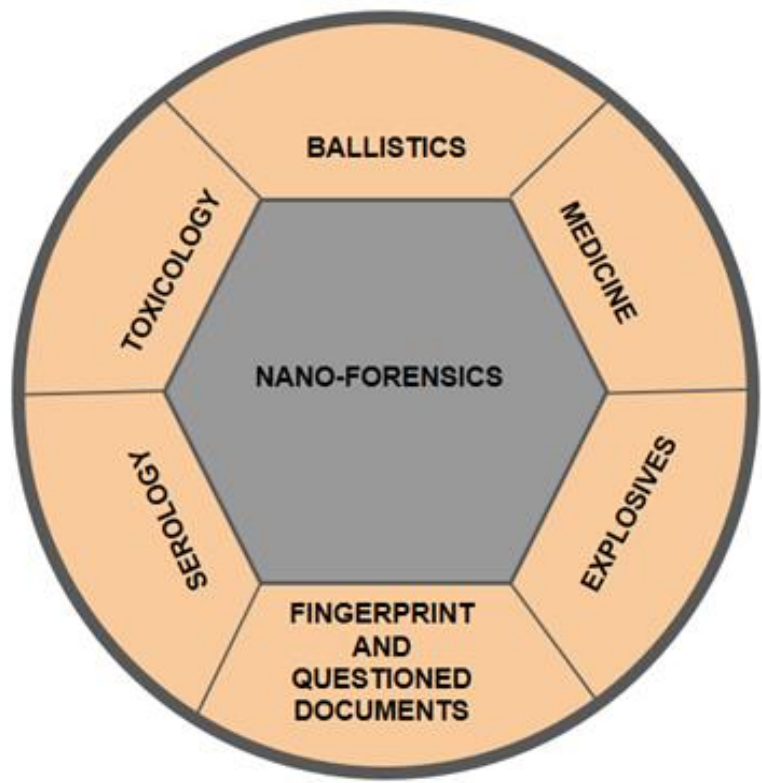

**Fig.1. Applications of nanotechnology in forensic science.**

2.1. Fingerprint Detection:

On the surface of the skin of fingertips, raised area of minute ridges are present known as the fingerprint ridges. These fingerprint ridges can form impressions when the fingertips come in contact with any surface or objects [17], [18]. The basic characteristics of fingerprints is that they do not change over the course of time (permanent) and every individual has unique fingerprints (individualistic, unique). This will aid them to get used as a tool for personal identification and also helps the investigators to reconstruct the sequence of events that led up to the crime [19]. Based on the visibility of the fingerprints, they have been classified as latent and patent fingerprints. Patent fingerprints are fingerprints that are basically visible to naked eye, whereas the latent fingerprints need to be visualized with the help of chemical, physical or optical methods as these fingerprints are less visible to the naked eye, this type of fingerprint is the most commonly encountered type at the crime scene [18]. Latent fingerprints are chemically composed of natural secretions which is mostly sweat from the Apocrine and Eccrine sweat glands that are present in the skin and contaminants from the environment such as cosmetics, dust, tobacco compounds etc.[20], [21]. Ninhydrin solution and iodine or benzoflavone spray (chemical methods) were used to develop the latent prints seen on porous surface. Whereas, superglue (cyanoacrylate esters) fuming is used for developing fingerprints on non-porous surfaces [22]. Various nano-based powders (physical methods) are also used for the development of latent fingerprints that are seen over the non-porous surface [17]. The fingerprint patterns are basically classified as Arch (~5%), loop (60-70%) and whorl (30-35%). These are again further sub-classified as plain arch and tented arch (arch); radial loop and ulnar loop (loop); plain whorl, central pocket loop, double loop, accidental loop and accidental whorl (whorl). These sub-classification of fingerprints are done based on the presence of delta, the direction of entry and exit of ridges, the flow of ridges and the number of delta [17], [23]. In the plain arch, the ridge enters from one side, forms a wave pattern and will exit through the opposite side, whereas in tented arch, the wave formed will rise to form a tent like shape. When it comes to the radial and ulnar loop the ridge flow happens from one side and exits through the same side. If the entry and exit of ridge is in the side of radial bone (thumb) it is known as the radial loop and if the entry and exit of ridge is in the side of ulnar bone (little finger) it is known as the ulnar loop. The plain whorl has two deltas and ridge should complete at one circuit [17]. If there is a formation of two loops and two deltas, they are categorized as double loop. If the pattern doesn't fit to any categories, they are classified as accidental pattern [17], [23]. Different types of fingerprint patterns are depicted in Fig.2.

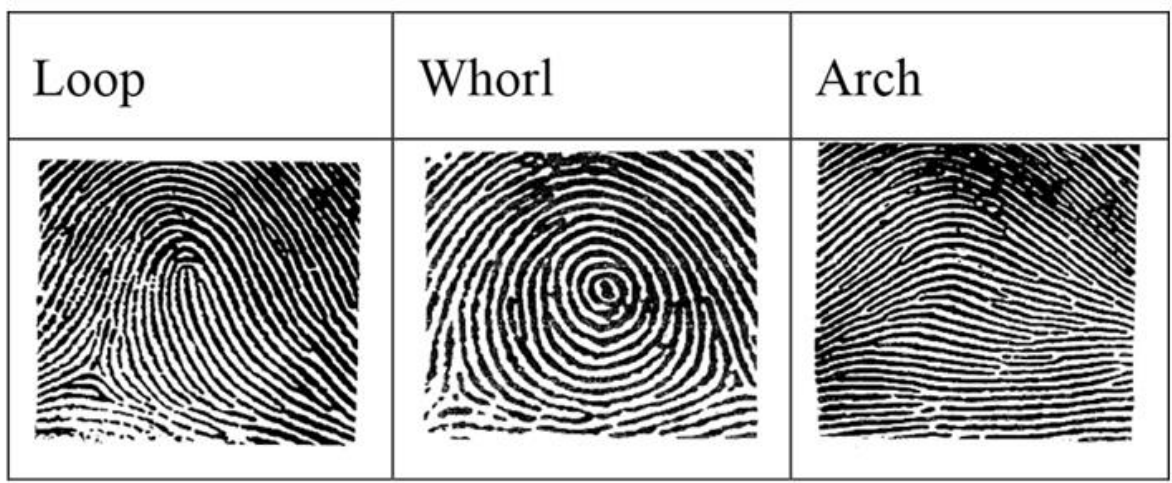

**Fig.2. Types of fingerprint patterns** [24].

2.2. Explosive Detection:

Detection of explosives is being difficult for law enforcement agencies each day as it is very expensive and intricate task due to various factors. Among many other explosives, TNT (Trinitrotoluene) is commonly found explosives in military applications. The potential of nanotechnology in detecting Trinitrotoluene (TNT), which is a common explosive compound is been studied widely. Pandya et al. developed a highly precise fluorescent probe for TNT detection using curcumin nanoparticles extracted from turmeric [25]. The probe was able to detect TNT at levels as low as 1 nm in aqueous solutions [26]. These studies suggest that nanotechnology can be employed to develop accurate methods for identifying TNT in various forensic samples. Compared to conventional methods, nanotechnology-based methods offer selectivity, sensitivity and portability. Forensic examinations of explosives include investigating post-explosion residue and detection of traces of explosives on the crime scene [27].

2.3. Toxicology:

The synthesis of designer drugs and psychotropic substances are generally done by modifying the existing drugs. For e.g. the two most frequently used classes of drugs are amphetamines and cathinone having quite similar structure. More sophisticated analysis is required for identification of such drugs and different toxic substances. Chemosensor is thus one such kind, for on spot detection and quantitative detection of drugs. The process of signal generation is mainly done by the changing properties of receptor-conjugated fluorescent dye, AFM or by indicator displacement. The technique called "nanoparticle-assisted NMR chemo sensing" is recently proposed for direct detection and identification of various classes of analytes. Here, the monolayer protected gold nanoparticles are used as self-organised receptor and vice versa. The nanoparticle receptor has ability to reorganize different metabolites, parts and form of drugs simultaneously [28]. Different sensors are also made for spot tests in cases of drug and alcohol. The visceral samples and various body fluids are collected and further detected for identifying various forms of drugs and toxic products in the body. Nanosensors in case of forensic toxicology helps to differentiate clonazepam drug from skeletal and blood samples by utilising gold nanoparticles [29].

2.4. Serology:

Nanotechnology serves as a very valuable forensic tool since it is unique to every individual, making it very useful for identifying people involved in a crime. Both DNA extraction and quantification are necessary steps in forensic caseworks. Extracted DNA can be used to make DNA profiles such as polymerase chain reaction (PCR) and short tandem repeats (STR), which are compared and used to link the suspect. Magnetic nanoparticles have emerged as a promising tool for high-purity DNA extraction from various biological samples due to their unique properties. These magnetic nanoparticles can be functionalized with several surface coatings to increase their affinity for DNA binding. Silica-coated magnetic nanoparticles have been widely used for DNA extraction as they offer a high surface area, biocompatibility, and good stability [30]. AFM, a rapidly developing technique is widely used for evaluating blood stains and thus providing useful information to legal medical experts regarding the (suspect/victim) forensic investigation and the case. The elasticity of blood can be easily detected by recording force-distance curve. In the area of nanotechnology, DNA could also be used as 'smart-glue' rather than a precise normal structure which doesn't provide high resolution but leads to an organized products [31].

2.5. Medicine:

In case of medicolegal investigation, time since death (TSD) plays a major role. The accurate estimation of the time helps the medicine expert and law to know the exact time and cause of death. There are certain parameters that are important to mention while investigating TSD such as algor mortis, rigor mortis, post-mortem hypostasis, changes in eye and observational changes in body fluids. Different body fluids like synovial fluid, spinal fluid, vitreous humor, aqueous humor etc. are also helpful in analysing TSD. As the vitreous humor remains unchanged for long in comparison to other body fluids, it can thus be used for estimating time since death. Therefore, the concentration of amino acid in vitreous humor can be detected by fluorescent nanoparticles and further quantification could be done by using flow cytometry which would lead to successful identification of TSD [25]. Fig.3. represents the schematic illustration of determination of time since death.

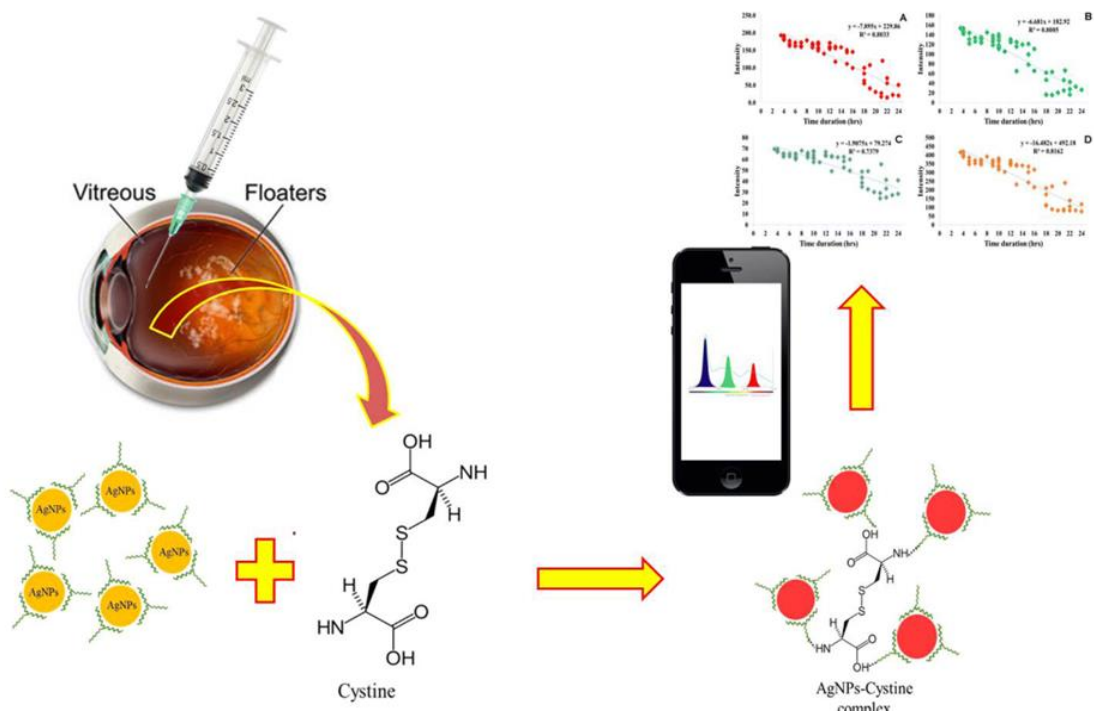

**Fig.3. Schematic representation of determining time since death from vitreous humor cysteine** [25]**.**

2.6. Ballistics:

In case of ballistics, the location of gunshot residue (GSR) as well as the elements that are present in GSR can be determined. Various advanced techniques such as SEM and SEM imaging are also used for the same. The perfect characterization of micrometric and sub micrometric features is critical in linking the ammunitions to a particular firearm. In order to obtain information on the shape and size of the GSR particles, the use of scanning probe microscope (SPM) has been marginally studied in forensic ballistics. Different striation marks that appear on the cartridge due to the mechanical parts like firing pin and breech face of the firearm after being fired can play a crucial role in assessing the origin of the cartridge. Less attempts have been made to characterize the morphology of the ammunition by nanotechnology-based techniques, however the use of nanotechnology in forensic science and investigation is increasing rapidly [32]–[36].

The instant detection of lead in GSR could be rapidly done on the crime scene, for this the GSR samples are collected from various evidential sources and fired ammunition which is further detected by the use of different metal nanoparticles such as Cu, Ag, Au etc. [37]. Cellulose and its derivatives, such as cellulose nanocrystals or cellulose nanofibers, due to its abundance and well-defined nano-scale structures are used in making lightweight materials for ballistics [38]. Carbon nanotubes (CNTs) are used in other areas of ballistics especially for the protection of light armoured and medium armoured military vehicles [39]. It has been also reported that due to the excellent strength, and high resistance to fracture, the polymer-reinforced CNTs becomes a potential material for the fabrication of ballistics armour [40]. Moreover, for fulfilling the current demands the use of graphene-based nanomaterials can also be used for the same [41]. Further, THz detectors of high efficiency can be made using carbon nanomaterials, particularly using CNTs as well as graphene-based materials because of their exciting optical, thermal and electronic properties [42].

## 3. Conclusion:

The significant role of nanomaterials in different sub-domains of forensic science such as fingerprint science, toxicology, ballistics, serology, medicine, explosive etc. has been reviewed. The nanotechnology-based upgradation of the techniques and methodologies used for investigation and imparting justice has also been discussed. It is concluded that the recent progress in the field of nanotechnology has led to great advancements in forensic science, which confirms the potential of this emerging field in serving justice. The frequent use and efficacy of nanotechnology has made it an advance technique in today's scientific world. In the modern scientific era, the combination of nanotechnology and forensic science will help in the exploration and generation of new investigating procedures.